\documentclass[12pt]{article}
\usepackage{epsf,latexsym}
\usepackage[all]{xy}
\usepackage{amsfonts,amssymb} 
\epsfverbosetrue
\newcommand{\de}{\hbox{\rm{d}}}

\newcommand{\bb}{\begin{eqnarray}}
\newcommand{\ee}{\end{eqnarray}}
\newcommand{\eee}{\nonumber\end{eqnarray}}
\newcommand{\qq}{\quad}

\begin{document}

\font\twelve=cmbx10 at 13pt
\font\eightrm=cmr8

\thispagestyle{empty}

\begin{center}
${}$
\vspace{3cm}

{\Large\textbf{Cosmological constant and lensing}} \\

\vspace{2cm}

{\large
Thomas Sch\"ucker\footnote{at Universit\'e de Provence,
schucker@cpt.univ-mrs.fr } (CPT\footnote{Centre de Physique
Th\'eorique\\\indent${}$\qq\qq CNRS--Luminy, Case
907\\\indent${}$\qq\qq F-13288 Marseille Cedex 9\\\indent${}$\qq
Unit\'e Mixte de Recherche (UMR 6207)
du CNRS et des Universit\'es Aix--Marseille 1 et 2\\
\indent${}$\qq et Sud
Toulon--Var, Laboratoire affili\'e \`a la FRUMAM (FR 2291)}) }

\vspace{3cm}

{\large\textbf{Abstract}}
\end{center}
The effect of the cosmological constant on the curvature of light due to an isolated spherical mass is recalculated without using the lens equation and compared to a lensing cluster.

\vspace{2cm}

\vskip 1truecm

\noindent CPT-2007/P93-2007\\
\vspace{2cm}

\section{Introduction}

Until september 2007, the official opinion had it that the deflection of light by an isolated spherical mass is independent of the cosmological constant. This was corrected in the beautiful work by  Rindler \& Ishak \cite{ri}. One month later, Ishak, Rindler et al. \cite{ir} showed that this dependence is not academic by exhibiting several lensing clusters, that do constrain the cosmological constant within today's bounds from the standard model of cosmology.

Two consecutive papers by Lake \cite{la} and Sereno \cite{se} confirm the  dependence. Sereno however finds a different expression for the deflection. Moreover he computes how the time delay is affected by the cosmological constant, which is particularly interesting in the light of the impressive progress \cite{fo} in the observation of such a time delay.

The aim of the present paper is to to give a detailed, self-contained account of how to compute the deflection in presence of a cosmological constant and only in terms of quantities observed in lensing clusters. In particular we do not use the lens equation. Our computations are to first order in Newton's constant and in the deflection angle. We do not neglect higher terms in the cosmological constant as they do contribute significantly in certain lensing clusters.

\section{The set up}

Consider a universe, which is empty except for one static, spherical, non-rotating mass $M$, the lens $L$. A source $S$, at rest with respect to the lens, emits photons, which are observed at nostra terra $T$ also assumed at rest. We neglect the masses of the source and of our local group. We use polar coordinates $(r,\theta ,\varphi )$ centered at the lens.  Because of spherical symmetry the photons' trajectory is in a plane that we take to be $\theta =\pi /2$. The angle $\varphi $ is measured with respect to the axis defined by the source, $\varphi _S=0$. With a cosmological constant, the gravitational field outside the mass $M$ is given by the Kottler metric,
\bb \de \tau ^2=B\,\de t^2-B^{-1} \de r^2-r^2\de\varphi ^2,\qq \theta =\pi /2,\qq 
B=1-\,\frac{2GM}{r} -{\textstyle\frac{1}{3}} \Lambda r^2.\ee
Our initial conditions for the photon are position $(r_S, \varphi _S=0)$ and velocity characterised by the coordinate angle $\epsilon_S:=r_S\,|\de\varphi /\de r\, (r_S)|$. Let $r_0$ be the peri-lens. We will suppose $2GM/r_0\ll 1$ and 
$\epsilon_S\ll 1$ and keep only terms linear in these two quantities. We will also suppose that $\Lambda r^2/3< 9/10$ to avoid the coordinate singularity at the equator of the de Sitter sphere.
The final conditions are here on earth $(r_T,\varphi _T)$ where the photons are detected together with their (coordinate) angle
$\epsilon_T:=r_T\,|\de\varphi /\de r\, (r_T)|$.

We are interested in situations where more than one  images of the source are seen. Because of the spherical symmetry there can be only one or two images depending on the distances of earth and source from the lens. There is however one exception, when earth, lens and source are aligned. Then $r_T=r_S$, $\varphi _T=\pi $ and we get an infinite number of images forming Einstein's ring.

We will concentrate on the non-aligned case and denote by $\epsilon '_S,\ r'_0,\ \epsilon '_T$ the parameters of the lower trajectory, figure 1.

\begin{center}
\begin{tabular}{c}

\xy
(0,0)*{}="L";
(60,0)*{}="S";
(-45,13)*{}="T";
(-45,13)*{\bullet};
(60,0)*{\bullet};
(62.5,-3)*{S};
(-47,9.5)*{T};
(-4,-2.5)*{L};
(0,0)*{\bullet};
{\ar (0,0)*{}; (70,0)*{}}; 
"S"; "T" **\crv{(0,20)};
"S"; "T" **\crv{(-10,-20)};
"L"; "T" **\dir{-}; 
"L"; (5,13)*{} **\dir{-}; 
"L"; (-2,-6.7)*{} **\dir{-}; 
(45,0)*{}; (47,4)*{} **\crv{(44,3)};
(43,0)*{}; (43.5,-4)*{} **\crv{(42,-2.5)};
(40,2.9)*{\epsilon _S};
(37,-2.5)*{\epsilon' _S};
(-33,9.6)*{}; (-31.5,14.5)*{} **\crv{(-30.5,12)};
(-29,8.4)*{}; (-31.5,3.3)*{} **\crv{(-29,5)};
(-25,11)*{\epsilon_T };
(-24,3)*{\epsilon' _T};
(5,0)*{}; (-4.9,1.4)*{} **\crv{(1,7)};
(6,6)*{r_0 };
(2,-4)*{r'_0};
(73,0)*{x};
(0,-14)*{};
(25,2.2)*{r_S};
(-15,6.7)*{r_T};
(-2.5,5.5)*{\varphi _T};

\endxy

\end{tabular}\linebreak\nopagebreak
{Figure 1: A double image}
\end{center}

Our tasks are:
\begin{itemize}\item
Integrate the photons' geodesics and compute $\varphi _T$ as a function of $r_S,\ \epsilon_S,\ r_T$ and $\varphi' _T$ as a function of $r_S,\ \epsilon'_S,\ r_T$. The surprising, but well known, fact is that in this calculation the cosmological constant drops out, which is not the case for massive geodesics.
\item 
Compute the coordinate distance $r_T$ as a function of $r_S,\ \epsilon_T,\ \epsilon'_T$, such that the two trajectories meet, $\varphi'_T=\varphi _T$.
\item
Compute the physically measured angles $\alpha _T$ and $\alpha' _T$ from the coordinate angles $\epsilon_T$ and $\epsilon'_T$. Here the cosmological constant re-enters the scene \cite{ri}.
\item
Compute the area distance $d_L$ of the lens as seen from earth  from the coordinate distance $r_T$.
\item
Compute the area distance $d_S$ of the source as seen from earth  from the coordinate distances $r_T$, $r_S$ and the angles $\alpha _T$ and $\alpha' _T$. 
Here the cosmological constant will also play a role.
\end{itemize}
Finally we will compare the theory to the observation of the lensing cluster SDSS J1004 +4112 with a quasar as source. The angles $\alpha _T$ and $\alpha' _T$ are measured as well as the mass of the lens and  the redshifts $z_L$ of the lens  and $z_S$ of the source 
\cite{sh,ot}.
\begin{itemize}\item
We will compute the area distances $d_L$ and $d_S$ from their redshifts $z_L$ and $z_S$ using the Hubble diagram for a flat 3-space with cosmological constant and 27 \% of matter.
\end{itemize}

\section{Integrating the geodesics}

We start with the list of the non-vanishing Christoffel symbols for the Kottler metric with $\theta =\pi /2$ and denote $':=\de/\de r$,
\bb {\Gamma ^t}_{tr}=B'/(2B), &
{\Gamma ^r}_{tt}=BB'/2, &
{\Gamma ^r}_{rr}=-B'/(2B),\\
{\Gamma ^r}_{\varphi \varphi }=-rB, &
{\Gamma ^\varphi }_{r\varphi }=1/r.\ee
The geodesic equations read:
\bb &&\ddot t+B'/B\,\dot t\dot r=0,\\
&&
\ddot r+{\textstyle\frac{1}{2}} BB'\dot t^2-{\textstyle\frac{1}{2}} BB'\dot r^2-rB\dot \varphi ^2=0,\\
&&
\ddot \varphi +2r^{-1}\dot r\dot
\varphi =0,
\ee
where we denote the affine parameter by $p$ and $\dot {}:=\de/\de p$. We immediately get three first integrals:
\bb \dot t=1/B,\qq r^2\dot\varphi =J,\qq
\dot r^2/B+J^2/r^2-1/B=-E.\ee
The last two come from invariance of the metric under rotations and time translations and the integration constants $J$ and $E$ have the meaning of angular momentum and energy per unit of mass. For the photon, $E=0$. Eliminating affine parameter and (coordinate) time we get:
\bb \frac{\de r}{\de \varphi }=\pm r\sqrt{r^2/J^2-B}.\label{dr}\ee
At the peri-lens, $\de r/\de \varphi (r_0)=0$ and therefore $J=r_0 B(r_0)^{-1/2}$. Substituting $J$ into equation (\ref{dr}),  the cosmological constant drops out and we have:
\bb\frac{\de \varphi }{\de r}\,= \pm\,
\frac{1}{r\sqrt{r^2/r_0^2-1}}\,\left[ 1-
\,\frac{2GM}{r}\,-\,\frac{2GM}{r_0}\,\frac{r}{r+r_0}\right]^{-1/2}.\ee
From now on we will omit terms of order $(GM/r_0)^2$, $\epsilon_S^2$ and $(GM/r_0)\epsilon_S$ and write equalities up to this order with a $\sim$ sign. For example:
\bb \epsilon_S=r_S|\de\varphi /\de r(r_S)|\sim
(r_S^2/r_0^2-1)^{-1/2}\sim r_0/r_S.\ee
Note that for the upper trajectory, $\de\varphi /\de r$ is negative for $r$ between $r_S$ and $r_0$, positive between
$r_0$ and $r_T$. Therefore
\bb \varphi _T=\int_{r_0}^{r_S}\left| \frac{\de \varphi }{\de r}\right|\,\de r+\int_{r_0}^{r_T}
\left| \frac{\de \varphi }{\de r}\right|\,\de r.
\ee
Using $\int x^{-1}(x^2-1)^{-1/2}\,\de x=-\arcsin 1/x$, $\int x^{-2}(x^2-1)^{-1/2}\,\de x=(x^2-1)^{1/2}/x,$
$\int (x+1)^{-1}(x^2-1)^{-1/2}\,\de x
=[(x-1)/(x+1)]^{1/2},$ we get to linear order:
\bb \varphi _T\sim\pi -\epsilon_S\left( 1+\,\frac{r_S}{r_T} \right) +\, \frac{4GM}{\epsilon_Sr_S}.\label{fiT}\ee 
For the lower trajectory the signs of $\de\varphi /\de r$ are opposite and we have
\bb \varphi' _T\sim\pi +\epsilon'_S\left( 1+\,\frac{r_S}{r_T} \right) -\, \frac{4GM}{\epsilon'_Sr_S}.\ee
We can trade $\epsilon_S$ for $\epsilon_T$ using
$r_0\sim\epsilon_Sr_S\sim\epsilon_Tr_T$ and likewise for the primed quantities.

For the two trajectories to meet, $\varphi _T=\varphi '_T$, we must have:
\bb \frac{r_T}{r_S} \,\sim\,\frac{4GM}{\epsilon_T\epsilon'_Tr_T}\,-1\label{meet}.\ee

\section{From coordinate angles to physical angles}

Let us compute the relation between the coordinate angle $\epsilon_T$ and the corresponding physical angle $\alpha _T$. Officially since october 1983, lengths are measured in terms of proper time of flight of photons, one nano-second $\sim$ 0.3 m $\sim$  1 foot. The coordinate time $\de t$ it takes the photon to travel from $(r_T,\varphi _T)$ to $(r_T-\de r,\varphi _T)$ is computed from $0=B(r_T)\,\de t^2-B(r_T)^{-1}\,\de r^2$, see figure 2.

\begin{center}
\begin{tabular}{c}

\xy
(0,-9)*{};
(0,15)*{};
(0,0)*{}="L";
(-45,13)*{}="T";
(-30,15)*{}="P";
(-32,9)*{}="A";
(0,0)*{\bullet};
(-45,13)*{\bullet};
(-49,14.5)*{T};
(-2,-3.5)*{L};
(-40,8.5)*{{\de \tau}};
(-27,12)*{\tilde{\de \tau}};
"L"; "A" **\dir{-}; 
(-19,2.3)*{r_T};
"P"; "T" **\dir{-}; 
"A"; "P" **\dir{~}; 
"A"; "T" **\dir{~}; 
(-30,15)*{\bf\cdot};
(-32,9)*{\bf\cdot};
(-30,15)*{\bf\cdot};
(-39,11)*{}; (-39,15)*{} **\crv{(-38,13)};
(-37,17.5)*{ \alpha _T};
(-39,15)*{\bf\cdot};
\endxy

\end{tabular}\linebreak\nopagebreak
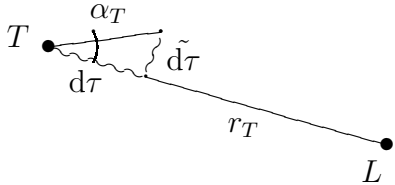
{Figure 2: Measuring the angle $\alpha _T$ in ns/ns}
\end{center}

Likewise the coordinate time $\tilde{\de t}$ it takes the photon to travel from $(r_T-\de r,\varphi _T)$ to $(r_T-\de r,\varphi _T-\de \varphi )$ is computed from $0=B(r_T)\,\tilde{\de t}^2-r_T^2\,\de \varphi ^2$. The we have,
\bb \tan \alpha _T=\,\frac{\tilde{\de\tau}}{\de\tau}\,=\,\frac{\tilde{\de t}\sqrt{B(r_T)}}{\de t \sqrt{B(r_T)}}\,=
\,\frac{r_T\,\de\varphi }{\de r/\sqrt{B(r_T)}}\,={\epsilon_T}{\sqrt{B(r_T)}}\sim
{\epsilon_T}{\sqrt{1-\Lambda r_T^2/3}}\sim\alpha _T.\ee
Similarly we get $\alpha '_T\sim
\epsilon'_T\sqrt{1-\Lambda r_T^2/3}.$
Rindler \& Ishak \cite{ri} compute the physical angle by means of the usual formula 
\bb\cos \alpha _T=\,\frac{(\de r,0)\cdot (\de r,\de\varphi )}{[(\de r,0)\cdot (\de r,0 )]^{1/2}[
(\de r,\de\varphi)\cdot (\de r,\de\varphi )]^{1/2}}\, ,\ee 
where the scalar product $\cdot$ comes from the negative of the spatial part of the Kottler metric. The two ways to compute the physical angle agree to all orders.

Finally we note that in Schwarzschild's solution and far out in the asymptotic region, physical and coordinate angles coincide. 

\section{From coordinate distances to area distances}

Imagine a standard candle radiating photons isotropically at the position of the lens. Its area distance $d_L$ as seen from the earth is defined by the relation between the infinitesimal solid angle $\de\Omega $ in which photons are radiated and the infinitesimal area $\de S$ of the light-sensitive plate  on which they are collected on earth, $\de S/\de\Omega = 4\pi d^2$. The isotropy of the standard candle is of course with respect to physically measured angles $\alpha $ which in this position do coincide with  the polar angles $\theta $ and $\varphi $, see figure 3.

\begin{center}
\begin{tabular}{c}

\xy
(0,20)*{};
(0,0)*{}="L";
(-45,13)*{}="T";
(-42.9,17)*{}="TT";
(0,0)*{\bullet};
(-45,13)*{\bullet};
(-47,9.5)*{T};
(-47,17.5)*{\de \ell_\parallel};
(-2,-3.5)*{L};
"L"; "T" **\dir{-}; 
"L"; "TT" **\dir{-}; 
"T"; "TT" **\dir{-}; 
(-13,8.8)*{\de\varphi_T};
(-29,5.3)*{r_T};
(-18.5,6)*{\bf\cdot};
(0,-10)*{};
\endxy

\end{tabular}\linebreak\nopagebreak
{Figure 3: Area distance of the lens as seen from earth}
\end{center}
\noindent
 The area is again measured by means of the time of flight of  test photons traveling along its edges. For the edge in the plane $\theta =\pi /2$ this coordinate time of flight is given by $0=B(r_T)\,{\de t}^2-r_T^2\,\de \varphi ^2$. The proper time is therefore $\de \tau = r_T \de\varphi $. For symmetry reasons we get the same proper time for  an edge orthogonal to the plane $\theta =\pi /2$ and consequently
 \bb d_L=r_T.\ee
 
For the area distance $d_S$ of the source as seen from the earth, the calculation is more involved. The isotropy of the standard candle in the plane $\theta =\pi /2$ is defined with respect to the physical angle $\alpha _S$ whose period is $2\pi\,B(r_S)^{1/2}$. The proper time of flight length of the infinitesimal edge at earth in this plane is $\de \ell_{\parallel}=r_T|\de\varphi _T|$, see figure 4.

\begin{center}
\begin{tabular}{c}

\xy
(0,24)*{};
(0,0)*{}="L";
(60,0)*{}="S";
(-45,13)*{}="T";
(-42.9,17)*{}="TT";
(0,0)*{\bullet};
(-45,13)*{\bullet};
(60,0)*{\bullet};
(62.5,-3)*{S};
(-47,9.5)*{T};
(-47,17.5)*{\de \ell_\parallel};
(-2,-3.5)*{L};
{\ar (0,0)*{}; (70,0)*{}}; 
"S"; "T" **\crv{(0,20)};
"S"; "TT" **\crv{(0,25)};
"L"; "T" **\dir{-}; 
"L"; "TT" **\dir{-}; 
"T"; "TT" **\dir{-}; 
(45,0)*{}; (47,4)*{} **\crv{(44,3)};
(40,2.9)*{\epsilon _S};
(5,0)*{}; (-4.9,1.4)*{} **\crv{(1,7)};
(73,0)*{x};
(25,-2.8)*{r_S};
(-14,8.8)*{\de\varphi _T};
(-29,5.3)*{r_T};
(5.5,5.5)*{\varphi _T};
(-18.5,6)*{\bf\cdot};
(27,10.3)*{\bf\cdot};
(29,14.3)*{\de \epsilon_S};
(0,-14)*{};
\endxy

\end{tabular}\linebreak\nopagebreak
{Figure 4: Area distance of the source as seen from  earth}
\end{center}

 Differentiating equation (\ref{fiT}) with respect to $\epsilon_S$ and using equation (\ref{meet}) we get
\bb \de\varphi _T&\sim&-\left( 1+\,\frac{r_S}{r_T}\,\right) \left(1+ \,\frac{\epsilon'_T}{\epsilon_T}\,\right)\,\de\epsilon_S
\sim-\left( 1+\,\frac{r_S}{r_T}\,\right) \left(1+ \,\frac{\alpha '_T}{\alpha _T}\,\right)\,\frac{\de\alpha _S}{\sqrt{B(r_S)}} \,\nonumber\\
&\sim&-\left( 1+\,\frac{r_S}{r_T}\,\right) \left(1+ \,\frac{\alpha '_T}{\alpha _T}\,\right)\,\frac{\de\alpha _S}{\sqrt{1-\Lambda r_S^2/3}} \,.
\ee
For an edge orthogonal to the plane $\theta =\pi /2$, the calculation is easier because of the axial symmetry around the axis lens - source. An infinitesimal rotation by a coordinate angle $\de\eta$ of period $2\pi $ results in an infinitesimal length at the earth of 
\bb \de\ell_{\perp}\sim r_T\sin(\pi -\varphi  _T)\,\de \eta\sim  r_T \left( 1+\,\frac{r_T}{r_S}\,\right) \,\frac{|\alpha _T-\alpha '_T|}{\sqrt{B(r_T)}}\,\de\eta .
\ee 
Replacing the coordinate angle $\de \eta$ by a physical angle we finally get the area distance from $\de S=\de\ell_{\parallel}\de\ell_{\perp}$:
\bb d_S\sim\,\frac{r_T+r_S}{\sqrt{1-\Lambda r_S^2/3}}\, \sqrt{\left(1+ \,\frac{\alpha '_T}{\alpha _T}\,\right)\,\frac{r_T}{r_S}\, \frac{|\alpha _T-\alpha '_T|}{\sqrt{1-\Lambda r_T^2/3}}}\,.\label{sing}
\ee
Note that a singularity occurs when earth, lens and source are aligned, $\alpha _T=\alpha '_T$. This comes from the focusing of the lens.

\section{From redshifts to area distances}

Here comes the shaky part of the reasoning. Indeed, as long as we do not have a solution of Einstein's equation interpolating between Kottler's and Friedmann's solutions, we do not know above what length scale the masses of the other galaxies and their expansion must be taken into account. Justified  only by this ignorance, we make the crude assumption that the measured redshifts are exclusively due to expansion, although we had put the  source and the earth at rest in the Kottler metric.  We will use the Hubble diagram from the standard model of cosmology with a flat 3-space, a cosmological constant $\Lambda = 1.5\cdot 10^{-52}\ {\rm m}^{-2}$, a Hubble parameter $H_0=2.5\cdot 10^{-18} \ {\rm s}^{-1}$ and 27 \% of matter in order to compute the area distances. A Runge-Kutta integration of Einstein's equation is used to obtain the scale factor $a(t)$.  For numerical convenience, we choose the initial condition $a_0=1.2\cdot 10^{26}$ m. Then the area distance $d$ follows from the redshift $z$:
\bb d(z)=\int_0^z\,\frac{\de a(t(\tilde z))/\de t}{a(t(\tilde z))} \,\de \tilde z\, ,\ee
where by abuse of notation we write $t(z)$ for the inverse function of $z(t):=1/a(t)-1$.
A rough fit to the numerical solution is given by \cite{fred}, 
\bb a(t)\sim a_0(pH_0t)^{1/p}, \qq p=0.69\,.\ee
It is good to  3 \% up to $z=2$ and yields
\bb d(z)\sim\,\frac{(z+1)^{1-p}-1}{(1-p)H_0}.\ee

Because of its singularity we cannot use the area distance $d_S$ (\ref{sing}) to estimate the  {\it position} of the source. The singularity is interpreted as magnification rather than an actual get-together. But then we need another assumption to estimate the position of the source.
The simplest assumption coming to mind is 
\bb d_S=\frac{r_T+r_S}{\sqrt{1-\Lambda r_S^2/3}}
\label{nosing}\ee 
which would be the correct area distance in absence of the lens. Of course one can argue that when we send the mass of the source to zero, we should send $\Lambda $  to zero at the same time, $d_S={r_T+r_S}$.

\section{SDSS J1004+4112}

Let us see how our assumptions compare to observation. Consider the lensing cluster of SDSS J1004+4112 and the quasar as source \cite{sh,ot}. As we have at least 4 images, the cluster cannot be spherically symmetric. We will again close our eyes  and consider only the images C and D with $\alpha _T=10''\ \pm 10\ \%$ and $\alpha' _T=5''\ \pm 10\ \%$. The mass of the cluster is $M=(1\pm 0.2)\cdot10^{44}$ kg. The cluster has a redshift of $z_L=0.68$ yielding $d_L=r_T=7.0\cdot 10^{25}$ m from the numerical integration. For the quasar we have $z_S=1.734$ and $d_S=13.7\cdot 10^{25}$ m,
which translates into $r_S=7.5\cdot10^{20}$ m with the singular area distance equation (\ref{sing}). However with this value the condition (\ref{meet}) \bb \frac{r_T}{r_S} \,\sim\,\frac{4GM}{\alpha _T\alpha '_Tr_T}\,(1-\Lambda r_T^2/3)-1\label{meet2}.\ee
for the two photon trajectories to meet cannot be satisfied whatever the value of $\Lambda $, zero or positive.
If we use the non-singular area distance 
(\ref{nosing}) we get $r_S=5.6\cdot10^{25}$ m and the condition for the trajectories to intersect can be met within the experimental error bars for the cluster mass $M$ and the angles $\alpha _T$ and $\alpha' _T$ only if the cosmological constant satisfies
\bb \Lambda > 0.81 \cdot 10^{-52}\ {\rm m}^{-2}.
\label{constr1}\ee
On the other hand, if we assume $d_S=r_T+r_S$ we get $r_S=6.7\cdot10^{25}$ m and the condition for the trajectories to intersect can be met within the error bars only if the cosmological constant satisfies
\bb \Lambda = (2.5\pm 1.5) \cdot 10^{-52}\ {\rm m}^{-2}.\label{constr2}\ee
In both cases we have not varied the cosmological constant in the Hubble diagram and it is encouraging that both constraints, (\ref{constr1}) and(\ref{constr2}) are compatible with the present observational bounds, $\Lambda = (1.5\pm 0.7) \cdot 10^{-52}\ {\rm m}^{-2}$, from the standard model of cosmology. Note that Ishak, Rindler et al. \cite{ir}, presumingly assuming $d_S=r_T+r_S$, find an upper bound from two other lensing clusters.

\section{Conclusions}

We agree with Rindler \& Ishak, gravitational lensing depends on the cosmological constant. Our formula (\ref{meet2}) for this dependence agrees with theirs \cite{ri} in the aligned case. We also agree with them \cite{ir}, this dependence is not negligible for certain clusters and further confrontation with observational data is necessary.

Of course the main problem remains to find a  solution of Einstein's equation that interpolates between a homogeneous family of static, curved Kottler solutions and the expanding, flat Friedmann solution. We hope that Rindler \& Ishak's ground breaking work will give new impetus to this fundamental problem.

Many years ago, dark energy was proposed as an alternative to the cosmological constant. And still, nobody has ever told us how dark energy modifies the Schwarzschild solution.
\vskip .5cm
\noindent {\bf Acknowledgements:} It is a pleasure to thank Christoph Stephan for lively discussions.


\begin{thebibliography}{10}

\bibitem{ri}
  W.~Rindler and M.~Ishak,
  ``The Contribution of the Cosmological Constant to the Relativistic Bending
  of Light Revisited,''
  Phys.\ Rev.\  D {\bf 76} (2007) 043006
  [arXiv:0709.2948 [astro-ph]].
\bibitem{ir}
  M.~Ishak, W.~Rindler, J.~Dossett, J.~Moldenhauer and C.~Allison,
  ``A New Independent Limit on the Cosmological Constant/Dark Energy from the
  Relativistic Bending of Light by Galaxies and Clusters of Galaxies,''
  arXiv:0710.4726 [astro-ph].
\bibitem{la}
  K.~Lake,
  ``More on the bending of light !,''
  arXiv:0711.0673 [gr-qc].
\bibitem{se}
  M.~Sereno,
  ``On the influence of the cosmological constant on gravitational lensing in
  small systems,''
  arXiv:0711.1802 [astro-ph].
\bibitem{fo}
  J.~Fohlmeister, C.~S.~Kochanek, E.~E.~Falco, C.~W.~Morgan and J.~Wambsganss,
  ``The Rewards of Patience: An 822 Day Time Delay in the Gravitational Lens
  SDSS J1004+4112,''
  arXiv:0710.1634 [astro-ph].
\bibitem{sh}
  K.~Sharon {\it et al.},
 ``Discovery of Multiply Imaged Galaxies behind the Cluster and Lensed Quasar
  SDSS J1004+4112,''
  Astrophys.\ J.\  {\bf 629} (2005) L73
  [arXiv:astro-ph/0507360].
\bibitem{ot}
  N.~Ota {\it et al.},
  ``Chandra Observations of SDSS~J1004+4112: Constraints on the Lensing Cluster
and Anomalous X-Ray Flux Ratios of the Quadruply Imaged Quasar,''
  Astrophys.\ J.\  {\bf 647} (2006) 215
  [arXiv:astro-ph/0601700].
\bibitem{fred}
  F.~Henry-Couannier, A.~Tilquin, A.~Ealet, A.~Bonissent, D.~Fouchez and C.~Tao,
  ``Negative energies and a constantly accelerating flat universe,''
  arXiv:gr-qc/0507065.

\end{thebibliography}
\end{document}